\documentclass[twocolumn,showpacs,preprintnumbers,amsmath,amssymb]{revtex4}
\usepackage{graphicx} 
\usepackage{dcolumn}  
\usepackage{bm}       
\begin{document} 
\title{Spin state transition in  LaCoO$_3$ by variational cluster approximation}
\author{R. Eder}
\affiliation{Karlsruher Institut f\"ur Technologie,
Institut f\"ur Festk\"orperphysik, 76021 Karlsruhe, Germany}
\date{\today}

\begin{abstract}
The variational cluster approximation 
is applied to the calculation of thermodynamical quantities
and single-particle spectra of LaCoO$_3$.
Trial self-energies and the numerical value of the Luttinger-Ward functional
are obtained by exact diagonalization of a CoO$_6$ cluster. 
The VCA correctly predicts LaCoO$_3$ as a paramagnetic insulator and
a gradual and relatively smooth increase of the occupation
of high-spin Co$^{3+}$ ions causes the temperature dependence
of entropy and magnetic susceptibility. The single particle spectral
function agrees well with experiment, the experimentally
observed temperature dependence of photoelectron spectra is reproduced
satisfactorily. Remaining discrepancies with experiment highlight
the importance of spin orbit coupling and local lattice relaxation.
\end{abstract} 
\maketitle
\section{Introduction}
LaCoO$_3$ has received considerable attention over the years
because it seems to undergo two electronic transitions or crossovers in the 
temperature range between 50 and 600 Kelvin\cite{Jonker,Raccah_Goodenough}.
The first crossover, usually referred to as the
spin state transition, can be seen most clearly
in the magnetic susceptibility $\chi$\cite{Bhide,Yamaguchi,Saitho}.
Below 50 Kelvin LaCoO$_3$ is nonmagnetic, $\chi \approx 0$,
indicating that all Co$^{3+}$ ions are in the low spin (LS) $^1A_{1g}$ or
$t_{2g}^6$ state realized for $d^6$ in cubic symmetry with
sufficiently large crystalline electric field (CEF).
Then $\chi$ rises sharply which indicates the thermal
excitation of states with nonzero spin and after a maximum around 100
Kelvin decreases again. 
In inelastic magnetic neutron scattering\cite{Asai} the
low frequency magnetic scattering intensity near $\Gamma$ shows a very similar 
temperature dependence as $\chi$. A pronounced anomaly
is also observed in the coefficient of thermal expansion\cite{Asai,Zobel},
the heat capacity  shows only a
weak anomaly at the spin state transition\cite{Stolen}.\\
Abbate {\em et al.}\cite{Abbate} found that
the valence band photoemission spectrum (PES) and O1s
X-ray absorbtion spectrum (XAS) show little or no
change across the spin state transition.
On the other hand, Haverkort {\em at al.}\cite{Haverkort} found a significant
temperature dependence of the Co-L$_{2,3}$
XAS and the X-ray magnetic circular dichroism spectrum (XMCD)
below 500 Kelvin. Thornton {\em et al.}\cite{Thornton_K_edge}
observed a temperature dependence of the
Co K-edge prepeak between 140 Kelvin and 800 Kelvin
and  Medarde {\em at al.}\cite{Medarde}
found that this temperature dependence sets in
at 50 Kelvin i.e. the onset of the spin state transition.\\
The nature of the spinful excited state which is responsible
for the spin state transition has been under debate for some
time. While it was proposed originally that this is
the high spin (HS) $^5T_{2g}$ (or $t_{2g}^4e_g^2$) excited state of 
the Co$^{3+}$ ion\cite{Raccah_Goodenough},
Korotin {\em et al.}\cite{Korotinetal} concluded from their LDA+U
calculation that this state rather is the intermediate spin (IS)
 $^3T_{1g}$ (or $t_{2g}^5e_g^1$) state.
Recently, however, experimental 
evidence has accumulated\cite{Noguchi,Ropka,Haverkort,Podlesnyak,Sundaram}
that it is really the HS state which is populated.
This leads to a certain puzzle in that a model calculation with an
$A_{1g}$ ground state and a $^5T_{2g}$ excited state with fixed
activation energy $\Delta=E(^5T_{2g})-E(^1A_{1g})$ cannot reproduce
the experimental $\chi(T)$ curve. If $\Delta$ is adjusted so as 
to reproduce the temperature where
$\chi$ starts to deviate from zero - $\approx$ 50 Kelvin -
the resulting maximum value of $\chi(T)$ near 100 Kelvin exceeds the
experimental value by a factor of $\approx 10$. The fit is much
better asuming an IS excited state, which has led 
some authors\cite{Saitho,Zobel}
to conclude that an IS state is responsible for the transition.\\
On the other hand, Haverkort {\em et al.} pointed out\cite{Haverkort}
that their XAS and XMCD spectra can only be explained by admixture of
a $^5T_{2g}$ excited state and concluded that the
activation energy $E(^5T_{2g})-E(^1A_{1g})$ is
{\em temperature dependent}, rising from
20 meV at 50 K to 80 meV at 700 K\cite{Haverkort} and leading to a
much slower increase of the population of HS ions with
temperature. A somewhat puzzling feature of this scenario is that in 
a situation  where the ground state is LS
an increase of the activation energy implies an {\em increase} of the 
CEF splitting $10Dq$ with temperature. 
The increase\cite{Radaelli_Cheong} of the 
Co-O bond length - the most important parameter
determining $10Dq$ - with temperature, however,
would result in exactly the opposite behaviour. The trend thus
cannot be explained in a single-ion-picture but is
a kind of `band effect'. 
Ky\^omen {\em et al.}\cite{Kyomen_1,Kyomen_2}, who deduced a very
similar temperature dependence of the activation energy
to reconcile magnetic susceptibility and specific heat data,
invoked a negative energy of mixing between low spin and high spin ions.\\
Another reason\cite{Haverkort} for
the smaller-than-expected value of $\chi$ observed in experiment
is spin-orbit coupling which splits the
$^5T_{2g}$ multiplet into three levels spanning an energy of 
$\approx 75\;m eV$\cite{Podlesnyak},
which is large compared to the temperature where the spin state transition
occurs. The lowest of these spin orbit-split states is threefold degenerate.
This low energy triplet - itself slightly split by
the trigonal distortion due to the orthorhombic
crystal structure of LaCoO$_3$ -
can be identified in electron spin resonance (ESR)\cite{Noguchi,Ropka} 
and inelastic neutron scattering\cite{Podlesnyak,Phelan}
experiments, whereby its $g$-factor of $\approx 3-3.5$ is clear proof
that it originates from a spin-orbit-split HS state rather
than from an IS state.\\
Haverkort  {\em et al}\cite{Haverkort} also found that to fit their
XAS and XMCD spectra with cluster calculations
they had to use a larger $10Dq$ for the $A_{1g}$ state
than for the $^5T_{2g}$ state. This hints at a participation of 
lattice degrees of freedom in that oxygen octahedra around
HS Co$^{3+}$ expand slightly so as to accomodate
the somewhat larger radius of the HS ion.
The emerging picture thus is a disordered mixture of LS and
HS states, whereby the lattice participates by an expansion of the 
CoO$_6$ octahedra around HS sites\cite{Haverkort}, which would
immediately explain the anomaly of the coefficient of 
thermal expansion\cite{Asai,Zobel}.
As pointed out by
Berggold {\em et al.}\cite{Berggold} this idea also nicely explains
an anomaly in the thermal conductivity $\kappa$
of LaCoO$_3$. At low temperatures the dominant contribution
to $\kappa$ comes from phonons and the expanded O$_6$ octahedra around HS
Co constitute randomly distributed lattice imperfections
which reduce the mean free path of the phonons. This leads to
a decrease of $\kappa$ at the onset of the spin state transition
and a minimum slightly below 200 Kelvin.\\
The idea of expanded O$_6$ octahedra around HS Co ions may resolve yet 
another puzzle, namely the result from inelastic magnetic
neutron scattering\cite{Asai,Phelan} that low energy spin correlations in  
LaCoO$_3$ are ferromagnetic rather than antiferromagnetic. 
This is surprising
in that the HS state has the configuration $t_{2g}^4e_g^2$ so that
the Goodenough-Kanamori rules would predict strong 
antiferromagnetic exchange interaction between two HS
Co ions on nearest neighbors. The expansion of the oxygen octahedra around
HS ions, however, would make the
formation of HS states on nearest neighbor Co ions energetically
unfavourable, so that this antiferromagnetic 
nearest-neighbor exchange may never have the chance to act. The ferromagnetic
spin correlations could then be due to `semiconductor version'
of the double exchange mechanism\cite{KhomskiiSawatzky}.\\
Strong experimental evidence against any appreciable occupation of
IS states is also provided by the EXAFS results of
Sundaram {\em et al.}\cite{Sundaram}. These authors ruled out
the existence of inequivalent Co-O bonds which would be practically
inevitable in the presence of IS states because the single electron
in the two $e_g$ orbitals would make these strongly Jahn-Teller active.\\
The second crossover in LaCoO$_3$ is frequently referred to as
a metal-insulator-transition. It can be seen most clearly in the specific 
heat where the raw data of  St{\o}len {\em et al.}\cite{Stolen}
show a sharp `spike' at 530 Kelvin even before subtraction of
the phonon background. Surprisingly for a metal-insulator-transition
the electrical conductivity $\sigma$ does not seem to show any noticeable
anomaly at this temperature.
Thornton {\em et al.}\cite{Thornton} found that at low temperatures
the electrical conductivity $\sigma$ shows a semiconductor-like
increase with temperature which can be fitted well by assuming an
activation energy of $\Delta=0.53 eV$ between 380 K and 520 K.
There is a broad plateau between 600 K and 800 K and only
above 800 Kelvin $\sigma$ decreases with temperature
as in a metal.
Bhide {\em et al.}\cite{Bhide} fitted the temperature dependence
of $\sigma$ with an activation energy between 0.1 eV - 0.2eV
for temperatures below 400 K. They found a plateau between
650 K and 1000 K and a decrease with temperature only above
1200 K.  Thornton {\em et al.}\cite{Thornton_1} inferred a 
`high-order semiconductor-to-metal
transition' between 385 Kelvin and 570 Kelvin from a study of
inflexion points in the $\sigma$ versus $T$ plot. \\
The magnetic susceptibility $\chi$ has a shallow maximum
near 600 Kelvin\cite{Bhide,Yamaguchi} 
whereas magnetic neutron scattering\cite{Asai} 
does not show a pronounced signature of the transition.
Abbate {\em et al.}\cite{Abbate} found a significant change in 
the O1s XAS spectra between 100 Kelvin and 570 Kelvin\cite{Abbate}
but the data of Thornton {\em et al.} (XAS at the Co K-edge)
show a similar change as the O1s XAS as the temperature changes from 
140 to 300 Kelvin so this change is not necessarily related
to the metal-insulator-transition.
Tokura {\em et al.} observed the filling of a gap-like structure
in the optical conductivity $\sigma(\omega)$, Richter {\em et al.}
did not observe a Fermi edge at temperatures above
the crossover\cite{RichterBaderBrodsky} in their
photoemission spectra. The evidence for
a true  metal-insulator-tansition thus is not really compelling and
in fact St{\o}len {\em et al.}\cite{Stolen} considered an entirely
different scenario, where the splitting of the $^5T_{2g}$ state
by spin orbit coupling plays a central rule. In this scenario,
the low temperature crossover is due to the thermal excitation of
the low-energy triplet, whereas the `metal-insulator-transition'
corresponds the population of the remaining components.\\
It has been argued that the crystal structure of LaCoO$_3$ may play 
a role as well. In the LDA+U calculations of
Korotin {\em et al.}\cite{Korotinetal} the structural change
with increasing temperature
is sufficient to induce a phase transition between
magnetic and nonmagnetic ground states.
Quite generally density functional calculations show a strong
sensitivity of ground state properties to structural
parameters\cite{Ravindran,Knizek}.
An experimental result result which directly shows the importance of 
lattice degrees of the lattice structure for the magnetic state of 
the Co$^{3+}$ ion is the ferromagnetism observed recently
in LaCoO$_3$ thin films under tensile strain\cite{Dirk_1}
Unlike bulk LaCoO$_3$ these thin films have temperature independent
photoelectron spectra\cite{Dirk_2}.\\
LaCoO$_3$ clearly is a difficult problem for any kind of
electronic structure calculation and has
been studied by various methods during the last years:
standard density functional thory\cite{Ravindran},
LDA+U or GGA+U\cite{Korotinetal,Knizek,Pandey_pes,Pandey_inverse,Hsu}
and dynamical mean-field theory\cite{Craco}.
As already mentioned
LDA band structure calculations incorrectly predict the material to
be a metal in the paramagnetic state for both, the ideal
perovskite structure and the true orthorombic structure\cite{Ravindran}.
Combined photoemission and 
bremsstrahlung isochromat spectroscopy (BIS) data\cite{Chainani}
indicate a gap in the electronic structure
although its precise magnitude is difficult to
pin down because the BIS spectrum shows a slow
and almost linear increase of intensity with increasing energy.
Together with the satellite structures observed
in valence band photoemission\cite{Saitho} this indicates the 
importance of electronic correlations and suggests that 
at low temperature the material is actually a correlated insulator.
From the above discussion it is moreover clear that a realistic
description of the temperature dependence of the photoelectron
spectra and magnetic susceptibility requires a 
correct description of the multiplet structure of the Co$^{3+}$ ion, 
and its interplay with the crystalline electric field. On the other hand
the relatively small gap indicates that covalency is strong
so that band effects obviously are important as well.
LaCoO$_3$ therefore appears as an interesting test case for the
variational cluster approximation proposed by Potthoff\cite{PotthoffI}.
This method generates trials self energies in a finite
cluster - an octahedral CoO$_6$ cluster in the present
implementation - so that the interplay between multiplet structure 
and crystal field splitting can be easily included. 
Being based on exact diagonalization rather than Quantum Monte Carlo
the VCA can access low temperatures as necessary for the
case of LaCoO$_3$. On the other hand, the present implementation
is based on an LCAO-fit to the band structure 
whose necessarily limited accuracy makes it hard
to quantitatively include the effects of changes of
the lattice. Therefore
all calculations were carried out for a rigid lattice,
which for simplicity was chosen to be the s.c. ideal
Perovskite structure. Bearing in mind the scenario inferred by
Haverkort {\em et al.} - an inhomogeneous
lattice distortion with CoO$_6$ octahedra expanding or contracting
{\em locally} in response to the spin state of the 
Co-ion - it is quite obvious that a quantitative
agreement with experiment cannot be expected for any calculation
for a rigid lattice. A quantitative discussion
of the temperature dependence of $\chi$ moreover would require to
include spin-orbit coupling which was omitted in the present study
to simplify the calculations.
Bearing this in mind we may expect the present calculation, with
a rigid lattice and no spin orbit coupling will reach
at best qualitative agreement with experiment. As will be shown below,
however, this goal is indeed achieved.
\section{Variational cluster approximation}
\begin{figure}
\includegraphics[width=\columnwidth]{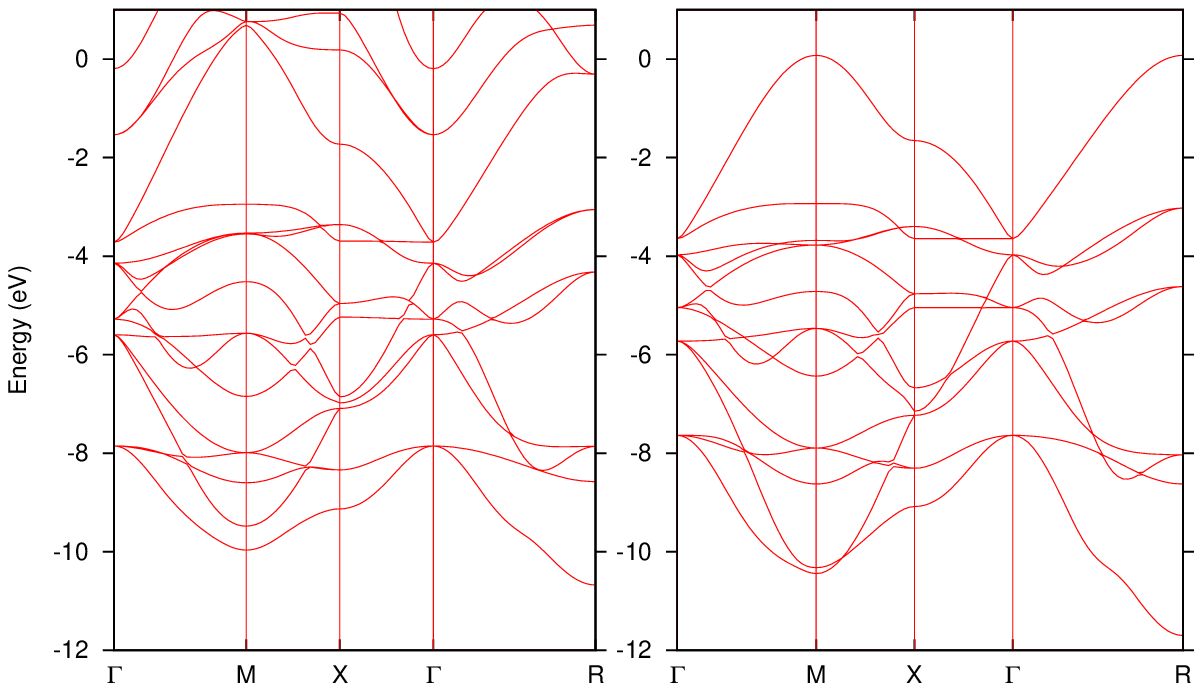}
\caption{\label{fig1}  (Color online)
LDA band structure (left) and LCAO-fit
(right) for LaCoO$_3$ in the ideal Perovskite structure.}
\end{figure}

The quantity which is subject to variation in the variational cluster
approximation (VCA) is the electronic self-energy $\Sigma(\omega)$.
More precisely the VCA seeks for the best approximation to the self-energy
$\Sigma(\omega)$ of a lattice system amongst the subset
of self-energies which can be represented as exact self-energies
of a given finite cluster.
The VCA is based on an expression for the  grand potential $\Omega$
of an interacting many-Fermion system derived by 
Luttinger and Ward\cite{LuttingerWard}. In a multi-band system
where the Green's function ${\bf G}({\bf k},\omega)$,
the noninteracting kinetic energy ${\bf t}({\bf k})$
and the self-energy ${\bf \Sigma}({\bf k},\omega)$
for given energy $\omega$ and momentum ${\bf k}$ are matrices of
dimension $2n\times 2n$, with $n$ the number of orbitals in the
unit cell, it reads\cite{Luttingertheorem}
\begin{eqnarray}
\Omega &=& -\frac{1}{\beta}\;\sum_{{\bf k},\nu}\; e^{\omega_\nu 0^+}
\ln\;det\;(-{\bf G}^{-1}({\bf k},\omega_\nu)\;) +
F[{\bf \Sigma}]
\label{ydef}
\end{eqnarray}
where $\omega_\nu=(2\nu+1)\pi/\beta$ with $\beta$ the inverse temperature
are the Fermionic Matsubara
frequencies,
\begin{equation}
{\bf G}^{-1}({\bf k},\omega) =\omega + \mu - {\bf t}({\bf k})
- {\bf \Sigma}({\bf k},\omega).
\label{gdef}
\end{equation}
with $\mu$ the chemical potential
and $F[{\bf \Sigma}]$ is the Legendre transform of the
Luttinger-Ward functional $\Phi[{\bf G}]$.
The latter is defined\cite{LuttingerWard} 
as the sum of all closed linked skeleton diagrams with the non-interacting
Green's functions replaced by the full Green's functions.
A nonperturbative derivation of a functional with the same properties
as $\Phi$ has been given by Potthoff\cite{Nonperturbative}.
Luttinger and Ward have shown that $\Omega$ is stationary with
with respect to variations of
${\bf\Sigma}$:
\begin{equation}
\frac{\partial \Omega}{\partial \Sigma_{ij}({\bf k},\omega_\nu)} = 0.
\label{stationary}
\end{equation}
but the crucial obstacle in exploiting this stationarity property 
in a variational scheme for 
${\bf \Sigma}$ is the evaluation of the functional $F[{\bf \Sigma}]$ for
a given `trial ${\bf \Sigma}$'. Potthoff's
solution\cite{PotthoffI} makes use of the fact that
$F[{\bf \Sigma}]$ has no explicit dependence on the single-particle 
terms of $H$ and therefore
is the same functional of ${\bf \Sigma}$ for any two systems
with the {\em same interaction part} of the Hamiltonian.
In the following only the Coulomb interaction 
within the $Co3d$-shell is taken
into account - which is a reasonable approximation.
Under this assumption $F[{\bf \Sigma}]$ then is the same functional for
the true perovskite lattice and for an array of identical but
disconnected octahedral CoO$_6$ clusters.
For given value of $\mu$ and $\beta$
one can therefore construct trial self-energies ${\bf \Sigma}(\omega)$
by exact diagonalization of a single CoO$_6$ cluster
and at the same time obtain the exact
numerical value of $F[{\bf \Sigma}]$ by simply reverting
the expression (\ref{ydef}).
Here the kinetic energy ${\tilde{\bf t}}$ of the CoO$_6$ cluster
has to be used in the Dyson equation (\ref{gdef}). Next, 
the pair $({\bf\Sigma},F[{\bf \Sigma}])$ can be used in (\ref{ydef}) for
the lattice system - which simply
amounts to replacing ${\tilde{\bf t}}$ by the kinetic energy
of the lattice, ${\bf t}({\bf k})$, in (\ref{gdef})
and performing the ${\bf k}$-summation - to obtain an
approximation for the grand potential of the lattice.
The variation of ${\bf \Sigma}(\omega)$ then is performed by varying the
single-electron parameters $\lambda_i$ - such as hybridization integrals or
site-energies - of the CoO$_6$ cluster.  The condition (\ref{stationary})
thus is replaced by the set of conditions
\begin{equation}
\frac{\partial \Omega}{\partial \lambda_i}=0.
\end{equation}
Potthoff has introduced the name `reference system' for the finite
cluster used to construct trial self-energies and computing the
Luttinger-Ward functional. In the present application - 
described in detail in  Refs. \cite{Eder_1} and \cite{Eder_2} - this is an 
octahedral CoO$_6$ cluster. Since it is known
that exact diagonalization of clusters comprising a single
transition metal ion and its nearest neighbor oxygens
gives excellent results for the ${\bf k}$-integrated photoelectron
spectra of many transition metal oxides\cite{FujimoriMinami,Degroot,Tanaka}
one may expect that the use of such a cluster as the reference
system is a reasonable choice. 
However, different implementations of the VCA 
have used quite different reference systems.
After being  proposed  by Potthoff\cite{PotthoffI}
(an excellent review covering many technical details
has been given by Senechal\cite{Senechal})
the VCA has been applied succesfully to 
one- and two-band Hubbard models\cite{Dahnken,Aichhorn},
to simplified models for to Fe pnictides\cite{Daghofer_1}
and to transition metal oxides with orbital degeneracy\cite{Daghofer}.\\
To obtain the single-electron Hamiltonian ${\bf t}({\bf k})$
an LDA band structure calculation for
LaCoO$_3$ was performed  using the Stuttgart LMTO-package. 
Thereby the ideal cubic perovskite structure with a Co-O
bond length of $1.91{\AA}$ was assumed. The density
of states is consistent with previous results
and actually quite similar to that obtained for the correct
rhombohedral structure\cite{Ravindran}. Next, an LCAO-fit was performed
to obtain a multi-orbital tight-binding parameterization
of the single-electron Hamiltonian ${\bf t}({\bf k})$.
The LCAO basis comprises O2s and O2p orbitals
at $-12.834\;eV$ and $0\;eV$, 
Co 4s and 3d orbitals at $19.436\;eV$ and $1.731\;eV$
and La 5p and 5d orbitals at $-9.264\;eV$ and $10.436\;eV$.
For the Co3d orbitals an additional $10Dq=0.848\;eV$ is obtained
from the fit.  All orbitals except O2p and Co3d only help to `polish' certain
portions of the band structure, but to obtain a good fit
they have to be included.
The energies of these `auxilliary orbitals' were not subject to the fit
only the respective two center integrals. These are listed in
Table \ref{tab1}. In general they refer only to nearest
neighbor bonds, but hybridization between second nearest neighbor
oxygen has also been used.
\begin{table}[h,t]
\begin{center}
\begin{tabular}{|c|rrrr|}
\hline
& Co-O & O-O & O-O & La-O \\
\hline

$(sp\sigma)$      &   -1.504 &   0.000 &   0.000 &   0.000 \\
$(pp\sigma)$      &    0.000 &   0.930 &   0.132 &   1.488 \\
$(pp\pi)$         &    0.000 &  -0.112 &   0.000 &  -0.289 \\
$(sd\sigma)$      &   -1.201 &   0.000 &   0.000 &   0.000 \\
$(pd\sigma)$      &    1.776 &   0.000 &   0.000 &  -0.879 \\
$(pd\pi)$         &   -0.975 &   0.000 &   0.000 &   0.296 \\
\hline
\end{tabular}
\caption{Two center integrals (in $eV$)
obtained by a LCAO fit to paramagnetic LDA band structure of LaCoO$_3$}.
\label{tab1}
\end{center}
\end{table}
Figure \ref{fig1} compares the actual LDA band structure and the
LCAO-fit - while the fit is not really excellent the
overall band structure is reproduced reasonably well.
The bands at the top of the figure which are absent in the LCAO
band structure originate from La$5d$ orbitals and would require to adjust
parameters such as La $(dd\sigma)$ - here they are put to zero because these
bands are not really interesting.
In the intervalls $-10\;eV \rightarrow -7\;eV$ and $-4\;eV \rightarrow 0\;eV$
there are bonding/antibonding
bands of mixed Co$3d$/O$2p$ character, in the intermediate energy range
there are essentially nonbonding
O$2p$ bands with very little Co$3d$ admixture.
The deviation from LDA seems large at the $R$ point, but it should be
noted that the band top at $M$ is also somewhat lower in the
LCAO-fit (the two band structures were alligned at $\Gamma$).
The total bandwith between the maximum at $M$ and the minimum at $R$
is $9.48\;eV$ for the LDA and $9.83\;eV$ for the LCAO-fit, i.e.
the difference is $4\%$ which seems tolerable.\\
The Coulomb interaction between Co 3d electrons is described by
standard atomic multiplet theory\cite{Slater,Griffith}. 
More precisely the Coulomb interaction within the $d$-shell can be written as
\begin{equation}
H_{1}=\sum_{\kappa_1,\kappa_2,\kappa_3,\kappa_4}
(\kappa_1 \kappa_2|g|\kappa_4 \kappa_3)\;\;
 d_{\kappa_1}^\dagger  d_{\kappa_2}^\dagger
d_{\kappa_3}^{} d_{\kappa_4}^{}.
\label{inter}
 \end{equation}
Here we have suppressed the site label $i$ and
$\kappa=(m,\sigma)$ where $m\in\{-2,-1,\dots 2 \}$ denotes the
$z$ component of orbital angular momentum.
The Coulomb matrix elements
$(\kappa_1 \kappa_2|g|\kappa_4 \kappa_3)$
are obtained by a multipole expansion of the
Coulomb interaction term $1/|{\bf r}-{\bf r}|$ and involve
Gaunt coefficients from the angular integrations
and the three Slater-Condon parameters 
$F^0$, $F^2$ and $F^4$ from the radial integrations.
The upper index thereby refers to the multipole order 
of the interaction and in a $d$-shell is limited to $\le4$ by the 
triangle condition. The somewhat lengthy complete expression
for the matrix elements is given e.g. in equations
(13-18)-(13-25) of the textbook by Slater\cite{Slater}.
$F^2$ and $F^4$ which describe higher multipole
interactions can be calculated from atomic Hartree-Fock wave
functions but $F^0$ is reduced substantially from its atomic value
by solid state screening and is treated as an adjustable parameter.
In the present work the values $F^0=8.376\;eV$, $F^2=10.64\;eV$ and
$F^4=6.804\;eV$ or, alternatively, the $3$ Racah-parameters, 
$A=7.62\;eV$, $B=0.14\;eV$, $C=0.54\;eV$ were used.
For the lowest mulpiplets the full theory as described by
(\ref{inter}) can be reduced
to a parameterization in terms of a Hubbard $U$ and
Hund's rule $J$, which parameters then can be expressed in
terms of the Slater-Condon parameters\cite{MarelSawatzky}.
As discussed in Ref. \cite{Eder_2}
we also need to specify the `bare' $d$-level energy $\tilde{\epsilon}_d^*$.
While the LCAO fit does give an energy $\epsilon_d$ for the Co3d level, this
contains a large contribution from the intra-$d$-shell
Coulomb interaction. 
Since we want to describe this Coulomb interaction
by adding the Coulomb Hamiltonian (\ref{inter}) to the 
LCAO-like single-particle
Hamiltonian we have to correct for this to avoid double counting.
For example, Kunes {\em et al.}\cite{Kunesetal} have estimated this
double counting correction as
$\tilde{\epsilon}_d^* = \epsilon_d - 9 U n_d$ where
$n_d$ is the average electron number/$d$-orbital.
With a $U$ of order $10\;eV$ this correction obviously is large.
Since first-principle calculations of screened interaction parameters
are a subtle issue, however,  $\tilde{\epsilon}_d^*$ was considered
as an adjustable parameter in the present work and set to be
$\tilde{\epsilon}_d^*=-46.4\;eV$.
$F^0$ and $\tilde{\epsilon}_d^*$ together essentially
determine the magnitude of the insulating
gap and the distance of the `satellite'  in the photoemission
spectrum from the valence band top.
With these values of the Racah parameters and $\tilde{\epsilon}_d^*$ and
using Table III of Ref. \cite{Eder_2} we obtain
the energy differences
$E(d^{n+1}) + E(d^{n-1}) - 2 E(d^n)$$=A-8B=6.5\;eV$
$E(d^{n+1}\underline{L})-E(d^n)$$=1.98\;eV$ which
are frequently referred to as the Hubbard $U$ and
the charge transfer energy $\Delta$.
Korotin {\em et al.}\cite{Korotinetal} obtained the value $U=7.5 \;eV$
by density functional calculations.\\
Finally, we discuss the CEF splitting 10Dq. It is obvious, that the
CEF is a crucial parameter for LaCoO$_3$ because it determines 
- amongst others - the relative energy of the $A_1g$ and $^5T_{2g}$ state
of the Co(3+) ion. One can not expect that the LDA calculation
and the LCAO-fit will produce a sufficiently accurate estimate
so as to reproduce energy scales of the order 100 Kelvin.
Therefore 10Dq was also treated as an adjustable parameter and
to get agreement with experiment the value 10Dq=0.72 eV was chosen, which 
still is rather close to the value of 0.848 eV obtained from the
LCAO fit.\\
Next we briefly comment on the technical problem of finding a stationary point
of $\Omega$ in a multi-dimensional parameter space.
As a first step, all but one parameter $\lambda_0$ are kept fixed
and  $\lambda_0$ is varied
until a value where $\partial \Omega / \partial \lambda_0=0$ 
is found. This means we are now on a surface in parameter space -
which we call the `$(\lambda_0)$-surface' -
where  $\partial \Omega / \partial \lambda_0=0$. 
It is advantageous to always choose $\lambda_0$ to be
the center of gravity of all orbital energies in the reference system 
because Aichhorn {\em et al.} have shown\cite{Aichhorn}, that optimization
of this parameter leads to a thermodynamically consistent
particle number.
Then a second parameter $\lambda_1$ is chosen and varied. 
In each step $\lambda_0$ is
recalculated to maintain $\partial \Omega / \partial
\lambda_0=0$ i.e. we walk along the $(\lambda_0)$-surface
in parameter space while varying $\lambda_1$.
The recalculation
of $\lambda_0$ can be done by means of the Newton-method,
thereby using the solution for the preceding value of $\lambda_1$
as initial guess for the next one. Variation of
$\lambda_1$ is continued until a value is found where
$\partial \Omega / \partial \lambda_1=0$. This means we have
now found a point of the  `$(\lambda_0,\lambda_1)$-surface ' 
in parameter space
which is defined by $\partial \Omega / \partial \lambda_0=0$ and
simultaneously $\partial \Omega / \partial \lambda_1=0$. Next, we choose a
third parameter, $\lambda_2$ and vary this again, walking
along the $(\lambda_0,\lambda_1)$-surface - 
the recalculation of $\lambda_0$ and $\lambda_1$ 
in each step is again done by the Newton method - until
we find a point where $\partial \Omega / \partial \lambda_2=0$
and so on. This method has the advantage that it is in principle
guaranteed to find a stationary point. Moreover
by doing a wider scan of one parameter one can
find different branches of the $\lambda$-surfaces which correspond
to different stationary points. \\
Finally we comment on the choice of the parameters to be
optimized. Using the notation of Ref. \cite{Eder_2} the $4$ parameters
$\epsilon_0$, $\epsilon_1$, $\epsilon_2$ and $V(e_g)$ were
varied. The values for the remaining parameters,
$\epsilon_3=1.4\;eV$ and $V(t_{2g})=2(pd\pi)$\cite{FujimoriMinami}
were kept fixed. It was checked that optimization of
more than $4$ parameters led to negligible change of $\Omega$
and very small changes in the single particle spectral function.
The reason for this `saturation' of $\Omega$ is the existence
of `nearly stationary' lines in parameter space as discussed
in detail in Ref. \cite{Eder_2}.
\section{Results}
\begin{figure}
\includegraphics[width=\columnwidth]{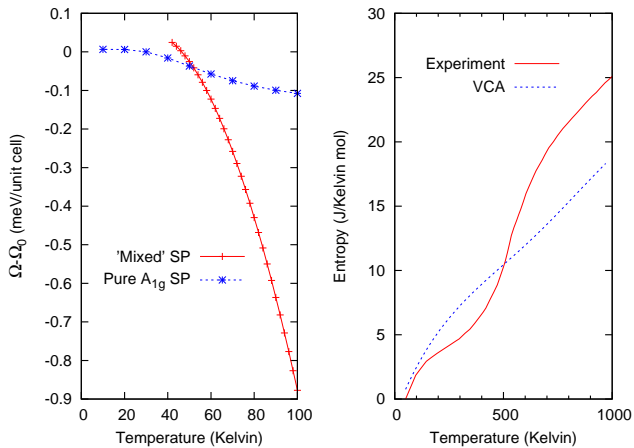}
\caption{\label{fig2}  (Color online)
Left: Grand canonical potential $\Omega$
for the two low energy stationary points
as a function of temperature. The value of $10Dq=0.72\;eV$ in the
lattice system.\\
Right: Entropy for the `mixed' SP as a function of temperature
compared to the experimental electronic entropy found by
St{\o}len {\em et al.}\cite{Stolen}.}
\end{figure}

A search in parameter space for stationary points (SP)  of $\Omega$
revealed that for most temperatures there are actually three different SP
corresponding to a d-shell occupation of $\approx 6$.
At low temperature the first one corresponds to
the reference system being in the pure A$_{1g}$ state,
the second SP corresponds to the reference system being in a thermal
mixture of an
A$_{1g}$  ground state and a $^5$T$_{2g}$ state at slightly higher energy, 
whereas the third SP corresponds to the reference system being in
the pure $^5$T$_{2g}$ state. For `reasonable'
parameters the third SP - corresponding to the HS state 
of the reference system - has an $\Omega$ that is substantially higher
than that of the first two SP, whence this SP will never be
realized. At low temperatures, on the other hand,
the A$_{1g}$-like SP and the `mixed' SP have very similar
values of $\Omega$ and for suitable choice of $10Dq$
{\em in the physical system} a crossover can be seen. To illustrate
this, Figure
\ref{fig2} shows $\Omega$ for the A$_{1g}$-like  and mixed SP
as a function of temperature for $10Dq=0.72\;eV$ (the value
of $10Dq$ obtained from the fit to the LMTO band structure
was $0.85\;eV$). Accordingly
this value of $10Dq$ was kept fixed for the rest of the calculation.
Much unlike the cases of NiO, CoO and MnO\cite{Eder_2}, the parameters
of the stationary points have a strong temperature dependence
in the case of LaCoO$_3$. Obviously this reflects
the subtle change with temperature of the electronic structure.\\
One can recognize in Figure \ref{fig2} that there is a crossing of the 
two $\Omega(T)$ curves
at $\approx 50\;$ Kelvin, and the finite difference in slope
implies a first order phase transition.
This way of describing the spin state transition is probably
an artefact of the mean-field-like description in the framework
of the VCA. The
latent heat for the transition would be  $T\Delta S=27.8\; J/mol$.
It should also be noted, that the $\Omega(T)$ curve of the
$A_{1g}$-like SP has an unphysical upward curvature above
$\approx 50 $ Kelvin. The lower SP,
however, does indeed have the correct downward curvature.\\
More interesting is the entropy $S(T)$ 
because this can be compared
directly to experiment. Figure \ref{fig2}
shows $S(T)$ from the VCA and the
experimental electronic entropy as extracted by 
St{\o}len {\em et al.}\cite{Stolen} from their specific heat
measurements. At low temperatures the agreement 
is not so bad but it is immediately obvious that the crossover
at $530\;$ Kelvin which is very pronounced in the entropy
is not reproduced at all by the present calculation. This will be discussed 
below.\\
It should also be mentioned that the IS (or $^3T_{1g}$) state has negligible
weight in the reference system even at the highest temperatures
studied. The reason is simply the fact - already noted by Haverkort
{\em et al.}\cite{Haverkort} - that the IS state never
comes even close to the ground state of the octahedral CoO$_6$
cluster. In that sense, the VCA complies with
`LS-HS scenario' supported by 
experiment.\cite{Noguchi,Ropka,Haverkort,Podlesnyak,Sundaram}\\
\begin{figure}
\includegraphics[width=\columnwidth]{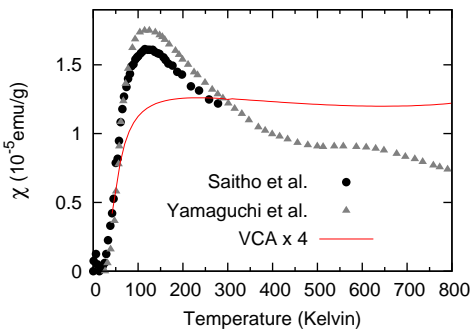}
\caption{\label{fig3}  (Color online)
Magnetic susceptibility of LaCoO$_3$ from Refs. \cite{Yamaguchi,Saitho}
and spin susceptibility obtained by the VCA. Note that the VCA result
is multiplied by a factor of $4$. The spin susceptibility in the
`A$_{1g}$-like SP, which is realized below 50 Kelvin, is essentially
zero.}
\end{figure}
Next, Figure \ref{fig3}
compares the spin susceptibility from the VCA calculation to the
experimental magnetic
susceptibility. While the overall behaviour is 
similar, the temperature where $\chi$ has its maximum does not agree with
experiment and, more importantly, the calculated values are
a factor $\approx 4$ too small. It is likely that the reason is the
rigid lattice used in the present calculation: in the actual material
the expansion of the O$_6$ octahedra around a Co-ion in the
high-spin state probably prevents HS ions from occupying
nearest neighbors, so that the strong antiferromagnetic superexchange
cannot act. In the VCA calculation this effect is absent,
whence the antiferromagnetic nearest-neighbor-exchange 
probably reduces the ferromagnetic
spin polarization induced by the magnetic field.\\
One may ask for the fraction of Co-ions being
in the HS state. It should be noted, that the VCA does not 
give that number for the physical system. 
The exact diagonalization of the reference system does give
the occupation numbers of the different eigenstates of the reference
system, but there is no justification for identifying these with the
occupation numbers in the physical system. On the other hand, if
the optimal self-energy for the lattice is realized in a cluster
where the the HS state has a certain weight $a$
it is reasonably plausible that the occupation of the HS state
in the physical system will not be differ completely from $a$. Thus,
we may consider the occupation of the HS state in the reference system
as a plausible estimate for the true HS occupation in the physical system.
 Figure \ref{fig4} compares this number to experimental
values. Most importantly  the increase of the
HS occupation with temperature is 
much weaker than for a system with fixed activation energy.
The estimates of Haverkort {\em et al} and Ky\^omen  {\em et al.}
are reasonably close and also the VCA gives a roughly correct
description although it obviously
underestimates the HS population. This is
another reason why the susceptibility computed by the
VCA is too small.
It is interesting, however, that the VCA gives the temperature dependence
of the HS occupation at least qualitatively correct as
it was carried out with a rigid lattice and therefore
includes `band effects' but no effects of the local lattice 
relaxation.\\
\begin{figure}
\includegraphics[width=\columnwidth]{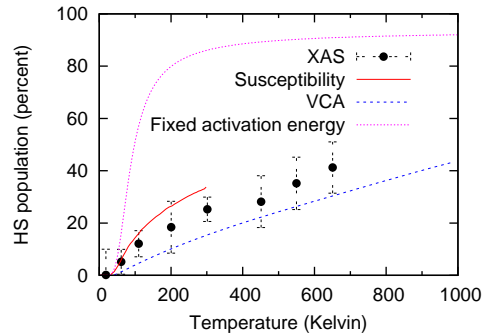}
\caption{\label{fig4}  (Color online)
Occupation of the $^5T_{2g}$ state in LaCoO$_3$ as inferred by 
Haverkort {\em et al.}\cite{Haverkort} from their XAS spectra
and by Ky\^omen  {\em et al.}\cite{Kyomen_2} from a fit to
the susceptibility and specific heat compared to
the $^5T_{2g}$ occupation in the reference system for the VCA
calculation. Also shown is the $^5T_{2g}$ occupation obtained
with a fixed activation energy of 250 Kelvin, which reproduces
the onset the spin state transition\cite{Saitho}.}
\end{figure}
Next we consider the occupation numbers of the various orbitals.
These can be obtained for the lattice system in the standard
way by integrating the
spectral function of the lattice system up to the chemical potential.
At 50 Kelvin, the occupation numbers/spin direction and atom
for the various orbitals are
$n(Co\;t_{2g})=2.956$, $n(Co\;e_g)=0.473$, $n(O2p)=2.823$.
It is immediately obvious from these numbers that there is
considerable charge transfer from Oxygen to the
$e_g$ orbitals of Cobalt.
\begin{figure}
\includegraphics[width=\columnwidth]{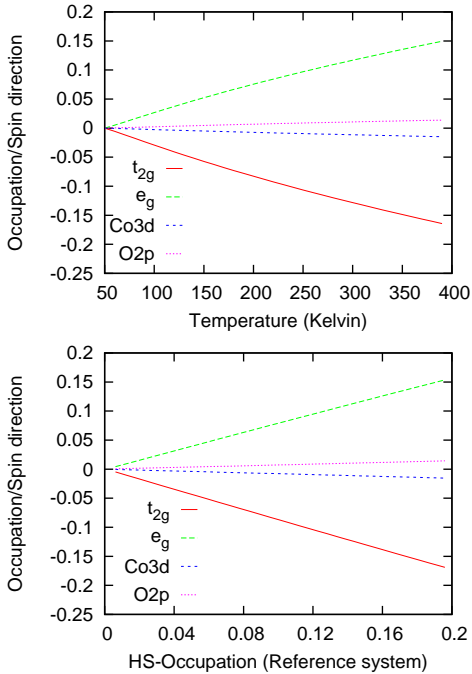}
\caption{\label{fig5}  (Color online)
Change of occupation numbers with temperature for the `mixed' SP.
The values at 50 kelvin have been subtracted off to make changes
more clearly visible. Also shown is the change of
occupation numbers plotted versus HS occupation in the reference
system.}
\end{figure}
As HS Co is admixed with increasing temperature, the occupation
numbers change, as can be seen in Figure \ref{fig5}.
The figure also shows the change of
occupation numbers plotted versus the
HS occupation {\em in the reference system}, i.e. the quantity
which is shown as `VCA' in Figure \ref{fig4}. The very accurate
linear dependence is an indication
that the HS occupation in the reference system is indeed
a very good estimate for the HS occupation in the lattice system.
The changes are as expected, with $n(Co\;e_g)$ increasing at the
expense of $n(Co\;t_{2g})$ and a slight net charge transfer
from Co to O. The nearly equal and opposite
change of  $n(Co\;t_{2g})$ and  $n(Co\;e_g)$is expected if 
HS $t_{2g}^4e_g^4$ are admixed to LS $t_{2g}^6$. Since
the $e_g$ orbitals hybridize with O by the stronger
$\sigma$ bonds and the $t_{2g}$ by the weaker $\pi$ bonds,
admixture of HS states will decrease the degree of covalency
hence the slight charge transfer back to Oxygen. More
precisely, each additional HS Co transfers $0.073$
electrons to the O2p bands.\\
Next, we consider the single-particle Green's function calculated with
the optimal self-energy. Figure \ref{fig6} shows the
${\bf k}$-integrated spectral function near the Fermi energy
as well as the combined PES and BIS spectra
by Chainani {\em at al.}\cite{Chainani}.
\begin{figure}
\includegraphics[width=\columnwidth]{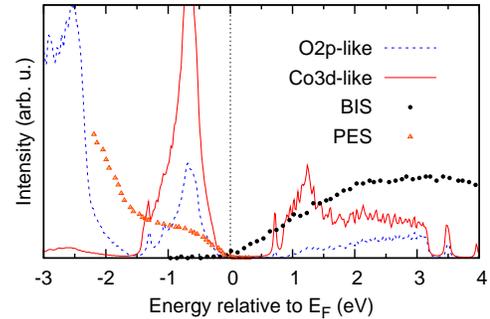}
\caption{\label{fig6}  (Color online)
Combined PES and BIS spectra (from Ref.\cite{Chainani})
compared to
the ${\bf k}$-integrated single particle spectral function obtained
from the VCA at 100 Kelvin. $\delta$-peaks are replaced by Lorentzians
with a width of 0.02 eV.}
\end{figure}
Most importantly, the VCA correctly describes LaCoO$_3$
as a paramagnetic insulator - there is a clear gap of $\approx 1\;eV$
in the spectrum. The BIS spectrum does not really show
a clear edge but a gradual increase so it is hard to
deduce a unique experimental gap value. 
The gap in the VCA spectrum of $\approx$ 1 eV is larger
than the gap values deduced from the
temperature dependence of the conductivity\cite{Bhide,Thornton}
which range from 0.1 eV to 0.53 eV (depending on temperature)
or from the optical conductivity, 0.1 eV\cite{Yamaguchi}.
On the other hand it should
be noted that the theoretical gap value does not have 
much real significance anyway - it is largely determined
by the adjustable parameters $U$ and $\Delta$. \\
More specific is the overall shape of the photoemission spectrum.
Figure \ref{fig7} compares the ${\bf k}$-integrated spectral function with the
experimental photoemission spectrum over a wider energy range. 
More precisely, we consider
the `On-off-difference', that means the difference of valence band
photoemission spectra taken with photon energies on ($63.5\;eV$)
and slightly off ($60.0\;eV$) 
the Co3p $\rightarrow$ 3d threshold, a procedure which is 
known\cite{Saitho} to emphasize the Co3d-derived features. Also shown 
is a photoemission
spectrum taken with a photon energy of $21.2\;eV$ where, due to the
larger photoionization cross section of the O2p 
orbital\cite{Eastman_Freeouf} at this energy,
mostly the oxygen derived states are visible.
The VCA reproduces the main features quite accurately:
the high intensity $d$-like peak at $-1\;eV$, the smaller
$d$-like peak at $-5\;eV$ and the broad `satellite' around
$-12\;eV$. Also, the $3$ O2p-like peaks are reproduced.
The fact that the peak at the top of the valence band has predominantly
Co3d character also comes out correctly.
By and large the VCA gives a reasonable description
of the electronic structure of LaCoO$_3$, at least on coarse
energy scales. \\
\begin{figure}
\includegraphics[width=\columnwidth]{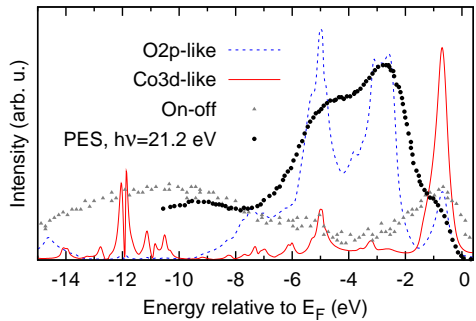}
\caption{\label{fig7}  (Color online)
Angle integrated valence band photoemission spectrum obtained
from the VCA for the `mixed' SP at 50 Kelvin compared to 
the 'on-off' spectrum and a photoemission
spectrum taken at $h\nu=21.2\;eV$ (Experimental data taken from 
Saitho {\em et al}\cite{Saitho}, Lorentzian broadening of the VCA specta:
0.1 eV).}
\end{figure}
An interesting feature of LaCoO$_3$ is the temperature dependence of
its photoelectron spectra and the VCA reproduces these at least
qualitatively. Figure \ref{fig8} shows the ${\bf k}$-integrated 
spectral function
at two different temperatures, 80 Kelvin and 570 Kelvin.
Also shown in the inset are experimental angle integrated photoemission
spectra by Abbate {\em et al.}. As one can see in experiment the prominent 
peak at the top of the valence band looses weight with increasing
temperature, which is a manifestation of the increasing
number of Co-ions in the high spin state. The VCA reproduces this effect
qualitatively, but higher temperatures are needed to
obtain a similar degree of spectral weight loss. As discussed above this
is simply due to the fact that the VCA underestimates the HS
occupation.\\
\begin{figure}
\includegraphics[width=\columnwidth]{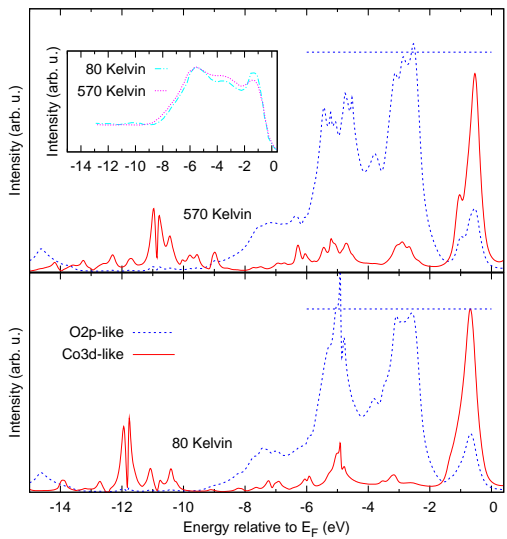}
\caption{\label{fig8}  (Color online)
Angle integrated valence band photoemisssion spectrum of LaCoO$_3$ 
obtained by VCA at 80 Kelvin (bottom)
and 570 Kelvin (top) (Lorentzian broadening 0.1 eV). 
The horizontal line is a guide to the
eye and corresponds to the same intensity in both spectra.
The inset shows experimental spectra at different temperatures
(Ref. \cite{Abbate}). }
\end{figure}
Next, we consider the unoccupied part of the spectrum and compare to
the Co K-edge spectra by Thornton {\em et al}\cite{Thornton_K_edge}.
Figure \ref{fig9} shows the spectral function at several temperatures
as well as the `prepeak' of the Co K-edge spectra - shifted in energy so as
to match the electron addition spectrum.
The VCA spectra show a peak at $\approx 0.7\;E eV$ above the Fermi energy
which increases in intensity as the temperature increases.
The experimental spectrum shows a similar change, i.e. the growing in
intensity of a low energy peak. Again, since the
VCA underestimates the HS population the growth of this
peak is probably underestimated. It should also be noted
that the O1s XAS spectra of 
Abbate {\em et al.}\cite{Abbate} show quite a similar bebaviour
as the Co K-edge spectra, namely the growth in spectral weight
of a low energy peak with increasing temperature. An apparent difference
between the results of Thornton {\em et al.} and 
Abbate {\em et al.} is that the Co K-edge spectra seem to show a rather
gradual change of the spectra with temperature, whereas the
O1s XAS spectra show little change at low temperature.
The VCA gives a very continuous and gradual change as would be
expected from the rather smooth increase of the HS population.\\
\begin{figure}
\includegraphics[width=\columnwidth]{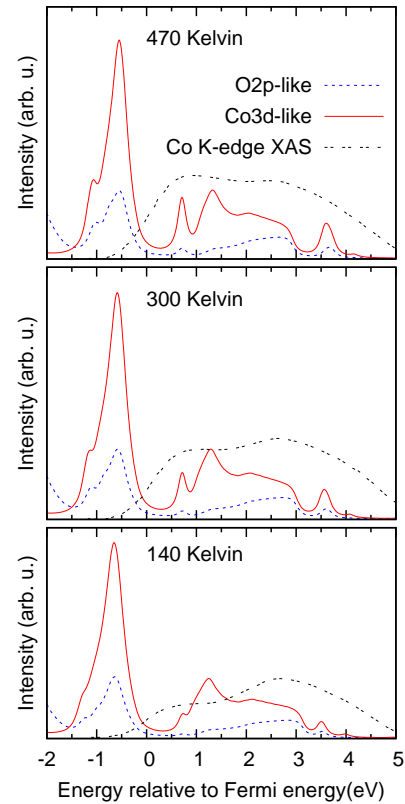}
\caption{\label{fig9}  (Color online)
${\bf k}$-integrated single particle spectral function
from VCA at different temperatures (Lorentzian broadening 0.1 eV)
compared to Co K-edge XAS spectra of LaCoO$_3$ 
(taken from Ref. \cite{Thornton_K_edge}). The XAS spectra have
been shifted downward by $\approx 7 \;keV$.}
\end{figure}
Lastly, we proceed to a comparison with other calculations
on LaCoO$_3$. Several authors have calculated the
density of states (DOS) for the low spin - or nonmagnetic - state
using LDA+U or GGA+U\cite{Korotinetal,Knizek,Pandey_pes,Pandey_inverse,Hsu}
and it may be interesting to compare the VCA to these calculations.
Figure \ref{fig10} shows the ${\bf k}$-integrated single particle spectrum
at the lowest temperature studied, 10 Kelvin. Whereas all spectra shown so far
corresponded to the `mixed SP' in Figure \ref{fig1}, 
this spectrum is calculated for the `pure A$_{1g}$' SP.
Despite this, one can see that the spectrum is nearly indistinguishable
from the other spectra, which shows that the phase transition
from the pure A$_{1g}$ SP to the mixed SP in Figure \ref{fig1}
has practically no influence on the spectrum. With the exception of the 
Co3d-like `satelite' at $\approx -12$ eV the spectrum is quite
consistent with the GGA+U calculation of 
Pandey {\em et al.}\cite{Pandey_pes}. Especially the respective
oxygen or Co character of the three prominent peaks agrees
reasonably well and these agree in turn with the photon-energy
dependence of the PES spectra\cite{Saitho}. The DOS
obtained by Hsu {\em et al.}\cite{Hsu} shows
three prominent peaks as well, but the characters do not match:
there, the topmost peak has predominant oxygen character, whereas
the lowermost peak has predominant Co character - this does not
agree with experiment. 
\begin{figure}
\includegraphics[width=\columnwidth]{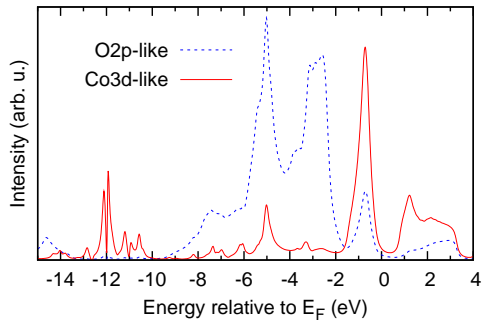}
\caption{\label{fig10}  (Color online)
${\bf k}$-integrated single particle spectrum obtained
from the VCA for the `A$_{1g}$-like' SP at 10 Kelvin,
Lorentzian broadening 0.1 eV).}
\end{figure}

\section{Discussion}
To summarize one may say that the VCA gives a reasonably
accurate description of some experimental results for LaCoO$_3$.
The insulating nature of the material is described correctly
and the photoelectron spectra agree with experiment in
quite some detail. The temperature dependence of the
magnetic susceptibility and the photoelectron spectra
is reproduced at least qualitatively. For all temperatures
studied only LS and HS states have appreciable weight
in the density matrix of the reference system, which means
that the VCA is consistent with the LS-HS scenario supported
by experiment. Thereby the population of the HS states
increases quite smoothly with an onset at 50 Kelvin which would be 
consistent with experiment as well.\\
The main deficiencies are the failure to reproduce the
crossover seen at 530 Kelvin in the entropy and susceptibility, 
the too small value of the magnetic susceptibility and the too 
slow increase of the HS population with temperature, which makes the
temperature dependence of all photoelectron spectra weaker than observed.
It should be noted that changing the values of $\Delta$ and/or $U$
so as to obtain e.g. a smaller insulating gap does not
change this. The above deficiencies are very probably not related to an
inappropriate choice of parameters.\\
The first reason for deviations is probably the neglect of 
spin-orbit-coupling in the $d$-shell. This leads to a splitting of the
$^5T_{2g}$ state into a three-fold, a five-fold and a seven-fold
degenerate state\cite{Noguchi,Ropka,Podlesnyak}
which span an energy of $75\;meV$. 
If one were to assume that the activation energy in the
reference system as obtained by the VCA
corresponds to the center of gravity of these split
states, the triplet would be appreciably lower and give a higher
susceptibility and HS occupation at low temperatures. Moreover,
St{\o}len {\em et al.}\cite{Stolen} have considered a scenario, where
the low temperature crossover corresponds to the population of
the low energy triplet and the high temperature transition to the population of
the remaining two components. If that were indeed the case,
a calculation without spin orbit coupling could never
reproduce the high temperature transition. Since spin-orbit
coupling is a single-particle term it can be included into
the VCA without any problem. On the other hand the z-component of the spin
is no longer a good quantum number if spin-orbit-coupling
is introduced, which increases the
size of matrices to be diagonalized or inverted and spin-orbit coupling
was neglected in the present study.\\
A second reason is the neglect of - or rather: the impossibility to treat -
the local lattice relaxations i.e. the expansion of O$_6$ octahedra
around HS Co ions. This implies than a HS ion `feels' a different
environment and that the actual activation energy has to be
modified by an elastic contribution. All of these effects are missed
in a calculation with a rigid lattice like the present one.
The local expansion of the O$_6$ octahedra may also have considerable
impact on the magnetic susceptibility. Namely based on the
Goodenough-Kanamori rules HS ions on nearest neighbors
would be expected to show strong antiferromagnetic exchange
via the two half-filled $e_g$ orbitals. On the other hand, for HS
ions on nearest neighbors it is clear that the respective
local lattice relaxations - expansion of the
 O$_6$ octahedra around HS ions - would interfere with each other,
so that HS occupation of nearest neighbors may be energetically
unfavourable and the antiferromagnetic superexchange may simply not
have the chance to act. This could explain the experimental 
result\cite{Asai,Phelan} that low energy spin correlations
are ferromagnetic rather than antiferromagnetic as well as the
surprising fact that thin films of LaCoO$_3$ under
tensile order ferromagnetically\cite{Dirk_1}. 
In a calculation with a rigid lattice this effect would be missed, 
so that the antiferromagnetic superexchange would reduce the 
spin susceptibility.\\
This shows that important physical effects had to be neglected
in the present calculation and a quantitative agreement with experiment
could not be expected. Still there is quite good qualitative agreement
which demonstrates the usefulness of the VCA to study correlated insulators.\\
I would like to thank K. P. Bohnen, D. Fuchs, M. Haverkort,
M. Potthoff and S. Schuppler for instructive discussions.
\section{Appendix: Calculation of the magnetic susceptibilitiy}
The magnetic susceptibility can be obtained from
$\chi=- \frac{\partial^2 \Omega}{\partial B^2}$. Thereby the
magnetic field $B$ is an additional single-particle-like parameter in the 
physical
system. The introduction of this parameter will change the
stationary point, that means the parameters of the reference system
become depedent on $B$. For the calculation of the derivative,
however, we do not need to solve the optimization
with applied $B$-field.\\
We denote by $\lambda_i$ the parameters of the reference system and by
$\bar{\lambda}_i$ the values at the stationary point for $B=0$.
Then we can write down the following expansion of $\Omega$
for small $B$:
\begin{eqnarray}
\Omega &=& \bar{\Omega} + 
\frac{1}{2}\sum_{i,j} 
\frac{\partial^2 \Omega}{\partial \lambda_i  \partial \lambda_j} 
(\lambda_i - \bar{\lambda}_i)(\lambda_j - \bar{\lambda}_j)\nonumber \\
&&\;\;\;\;\;\;+\sum_i
\frac{\partial^2 \Omega}{\partial \lambda_i  \partial B}
B (\lambda_i - \bar{\lambda}_i) 
+ \frac{B^2}{2} \frac{\partial^2 \Omega}{\partial B^2}
\label{eq1}
\end{eqnarray}
All derivatives in this equation can in principle be
obtained numerically.
Also it has been used that in a paramagnetic state
\[
\frac{\partial \Omega}{\partial B}=0.
\]
Taking $B$ small but finite and demanding that
$ \frac{\partial \Omega}{\partial \lambda_i}=0$ we obtain
\[
\sum_{j} 
\frac{\partial^2 \Omega}{\partial \lambda_i  \partial \lambda_j} 
(\lambda_j - \bar{\lambda}_j) = - 
\frac{\partial^2 \Omega}{\partial \lambda_i  \partial B} B
\]
We now differentiate with respect to $B$ and set $B=0$ to obtain
\[
\sum_{j} 
\frac{\partial^2 \Omega}{\partial \lambda_i  \partial \lambda_j} 
\frac{ \partial \lambda_j}{\partial B} = -
\frac{\partial^2 \Omega}{\partial \lambda_i  \partial B} 
\]
which is an equation for
the derivatives $\frac{ \partial \lambda_j}{\partial B}$.
We assume this to be solved and thus the 
$\frac{ \partial \lambda_j}{\partial B}$ to be known.
Inserting now
$\lambda_i - \bar{\lambda}_i=B\cdot \frac{ \partial \lambda_j}{\partial B}$
into equation (\ref{eq1}) we obtain
\begin{equation}
\Omega = \bar{\Omega} + 
\frac{B^2}{2}\left[ 
\frac{\partial^2 \Omega}{\partial B^2} 
+\sum_i
\frac{\partial^2 \Omega}{\partial \lambda_i  \partial B}
\frac{ \partial \lambda_j}{\partial B}\right]
\label{eq2}
\end{equation}
from which the susceptibility is found as
\[
\chi = -\frac{\partial^2 \Omega}{\partial B^2} 
-\sum_i
\frac{\partial^2 \Omega}{\partial \lambda_i  \partial B}
\frac{ \partial \lambda_j}{\partial B}
\]
In the presence of a magnetic field all single electron
parameters of the reference system have to be taken spin dependent,
i.e. a hopping integral $t\rightarrow (t_\uparrow, t_\downarrow)$.
It is then easy to see that the mixed second derivatives
$\frac{\partial^2 \Omega}{\partial \lambda_i  \partial B}$
are different from zero only for those $\lambda_i$ which are
odd under sign change of the spin, that means
quantities like $(t_\uparrow- t_\downarrow)$. At a nonmagnetic
SP these are zero so the derivatives can be evaluated
right at the SP.

\end{document}